\documentclass[12pt, draftclsnofoot, onecolumn]{IEEEtran}
\makeatletter
\def\endthebibliography{%
  \def\@noitemerr{\@latex@warning{Empty `thebibliography' environment}}%
  \endlist
}
\makeatother
\ifCLASSINFOpdf
   \usepackage[pdftex]{graphicx}
\else
 \usepackage[dvips]{graphicx}
\fi

\usepackage{graphicx}
\usepackage{amsmath}
\usepackage{amssymb}
\usepackage[caption=false]{subfig}
\usepackage[noadjust]{cite}
\usepackage{float}
\usepackage{algorithm}
\usepackage{bibentry}
\usepackage{balance}
\usepackage{algorithm}
\usepackage{algorithmicx}
\usepackage{algpseudocode}
\usepackage{xcolor}
\usepackage{balance}
\usepackage{enumitem}
\usepackage{multirow}

\begin{document}
\renewenvironment{description}[1][12pt]
  {\list{}{\labelwidth=0pt \leftmargin=#1
   \let\makelabel\descriptionlabel}}
  {\endlist}
 \newcommand\norm[1]{\left\lVert#1\right\rVert}
 \setlength{\abovedisplayskip}{0.1cm}
\setlength{\belowdisplayskip}{0.1cm}

\title{Rate Splitting for Multi-group Multicasting with a Common Message}

\author{Ahmet Zahid Yalcin,~\IEEEmembership{}
        Melda Yuksel,~\IEEEmembership{Senior Member, IEEE},~and \\
        Bruno Clerckx,~\IEEEmembership{Senior Member, IEEE}
}

\maketitle
\begin{abstract}

In this letter, precoding for max-min fairness (MMF) for multi-group multicasting with a common message is studied. The MMF problem is converted into a weighted mean square error minimization problem. A rate-splitting solution is proposed. In rate-splitting, multicast messages for each group are divided into private and common parts, and these common parts, together with the original common message are combined as a \emph{super common} message. This super common message is superposed on or concatenated to the private multicast data vector, or it is transmitted via a mixed scheme. Simulations show that RS demonstrates significant gains especially in overloaded systems.

\end{abstract}


\begin{IEEEkeywords}
Common message, max-min fairness, multi-group multicasting, multiple-input multiple-output, precoding design, rate-splitting.
\end{IEEEkeywords}

\IEEEpeerreviewmaketitle

\section{Introduction}
In emerging applications such as intelligent transportation and public warning systems, on-demand video, and applications for machine type communications, groups of users require the same messages. To utilize system resources efficiently, these messages should be precoded in a multicast fashion \cite{Sidiropoulos2006_TransmitBeamforming, Jindal2006_CapacityLimitsofMultipleAnt, Abdelkader2010_MultipleAntennaMulticasting}, specially designed for each network topology. In state-of-the-art, long term evolution (LTE) systems, multicasting is achieved via the enhanced multimedia broadcast multicast service (eMBMS) interface \cite{multicast_2}. For fifth generation (5G) and beyond, it is essential to advance these interfaces and the physical layer techniques to account for multigroup multicasting as well.



Multi-group multicasting is first studied in \cite{Silva2009_LinearTransmitBeamforming}. Two optimization problems for multi-group multicasting, minimizing transmission power under quality of service (QoS) constraints, and maximizing fair rate under a total power constraint, are solved in \cite{Karidipis2008_QoSMaxMinFairTransmit}. This work is extended for per antenna power constraints in \cite{Christopoulos2014_WeightedFair}. Rate-splitting (RS) for multi-group multicasting is proposed in  \cite{Joudeh2017_RateSplittingforMaxMin} for better inter-group interference management. Multi-group multicasting with a common message is studied in \cite{Yalcin2019_MGMC_withCommon}, and superposition coding is suggested as an efficient method to transmit the common message. RS in a multiple input single output broadcast channel with a common message is investigated in \cite{Mao2019_RSforMultiAnt}. It shown that the successive interference cancellation (SIC) architecture needed to separate the common message from the unicast streams can be used more efficiently by adopting an RS based transmission strategy that encodes the common message and part of the unicast messages into a super common message.

In this letter, we build upon the benefits demonstrated in \cite{Joudeh2017_RateSplittingforMaxMin,Yalcin2019_MGMC_withCommon,Mao2019_RSforMultiAnt} and look at RS in multi-group multicasting with a common message. Compared to \cite{Joudeh2017_RateSplittingforMaxMin}, this letter considers the presence of a common message. Compared to \cite{Yalcin2019_MGMC_withCommon}, this letter considers an RS strategy. Finally, when compared to \cite{Mao2019_RSforMultiAnt}, this letter considers multi-group multicasting.
Moreover, in this letter, precoders are designed based on RS and superposition ideas and three different schemes are compared: (i)~the \emph{super common} message is superposed on the multicast messages, (ii) the \emph{super common} message is concatenated to the multicast message vector, and (iii) the \emph{super common} message is transmitted via a combination of both. The results show that RS introduces significant gains in overloaded systems; i.e. when the total number of users is larger than the number of transmit antennas.

The rest of the letter is organized as follows. The system model is described in Section 
\ref{sec:systemmodel}. The optimization problem is defined in Section \ref{sec:Probdef}. Simulation results are presented in Section \ref{sec:sim}, and the letter is concluded in Section \ref{sec:conc}.

\section{System Model}\label{sec:systemmodel}
We consider a wireless system comprising of a single base station equipped with $M$ antennas and $N$ single-antenna receivers indexed by the set $\mathcal{N} \triangleq  \{1, . . . , N\}$. Receivers are grouped into the $K$ multicast groups $\mathcal{G}_1,\ldots,\mathcal{G}_K$, where $\mathcal{G}_k$ is the set of receivers belonging to the $k$th group, $k \in \mathcal{K}$, $\mathcal{K}  \triangleq \{1,...,K\}$, and $1 \leq K \leq N$. It is assumed that each receiver belongs to exactly one group. Thus $\bigcup_{k\in \mathcal{K}} \mathcal{G}_k = \mathcal{N}$ and $\mathcal{G}_k \bigcap \mathcal{G}_j = \emptyset $, $\forall k,j \in \mathcal{K}$ and $k \neq j$. Denoting the size of the $k$th group by $G_k = |\mathcal{G}_k|$, it is assumed without loss of generality that group sizes are in an ascending order, i.e. $G_1 \leq G_2 \leq \ldots G_K$.
To map users to their respective groups, we define $\mu : \mathcal{N}\rightarrow \mathcal{K}$ such that $\mu(n) = k$ for all $n \in \mathcal{G}_k$. 

The base station wants to transmit a system-wide common message $S_c$ intended for all $N$ users and $K$ multicast messages $S_1 ,\ldots , S_K$ intended for different groups. Using RS, the multicast message $S_k$ intended for group $k$ is split into a common part $S_{c,k}$ and a private part $S_{p,k}$, $\forall k \in \mathcal{K}$. The common parts of the multicast messages $S_{c,1}, \ldots, S_{c,K}$ are encoded along with the system-wide common message $S_c$ as a super-common message $S_0 = S_c \bigcup S_{c,1} \bigcup \ldots \bigcup S_{c,K}$. This super common message is required to be decoded by all users. Note that $S_0$ includes not only the common message, but parts of the multicast messages intended for different groups. The super-common message $S_0$ and the private parts of the multicast messages $S_{p,1} ,\ldots, S_{p,K}$ are independently encoded into $s_0, s_{p,1},\ldots,s_{p,K}$.

\begin{description}
\item[Superposition Coding Scheme ($\mathsf{SC}$): ] 
In the first signal model, the base station employs a 2-layer superposition coding scheme, where the base layer carries the super common message and the enhancement layer carries the private parts of the multicast data. The input data vector is denoted as $\mathbf{s}^\mathsf{SC} =\mathbf{s}_0^\mathsf{SC} + \mathbf{s}_p^\mathsf{SC}$, where $\mathbf{s}_0^\mathsf{SC} = {[s_0,\ldots,s_0]}^T$ $\in \mathbb{C}^{K}$ and $\mathbf{s}_p^\mathsf{SC} = {[s_{p,1},\ldots, s_{p,K}]}^T$ $\in \mathbb{C}^{K }$. We assume $s_0$ and all $s_{p,k}$ are independent and
$\mathbb{E}\{s_0 s_0^{\ast}\} = \alpha$ and  $\mathbb{E}\{{s}_{p,k}{s}_{p,k}^{\ast}\} = \bar{\alpha}$. Here, $\bar{\alpha} = 1-\alpha$ and $\alpha$ is the ratio of power allocated to the super-common data. The input data vector $\mathbf{s}^\mathsf{SC} $ is linearly processed by a precoder matrix $\mathbf{P}^\mathsf{SC}= [\mathbf{p}_1^\mathsf{SC},\ldots,\mathbf{p}_{K}^\mathsf{SC}]$. Each precoding vector $\mathbf{p}_k^\mathsf{SC}\in \mathbb{C}^{M}$ is of size $M \times 1$.

\item[Concatenation Coding Scheme ($\mathsf{CC}$):] 
In this scheme, the input data vector is defined as $\mathbf{s}^\mathsf{CC}= {[s_0, s_{p,1},\ldots,s_{p,K}]}^T \in  \mathbb{C}^{K+1}$,  where $s_0$ and $s_{p,k}$ are the same as in the $\mathsf{SC}$ model. We assume $s_0$ and all $s_{p,k}$ are independent and $\mathbb{E}\{|s_0|^2\} \,{=}\, 1$ and $\mathbb{E}\{s_{p,k} s_{p,k}^*\} = 1$. The input data vector $\mathbf{s}^\mathsf{CC}$ is linearly processed by a precoder matrix $\mathbf{P}^\mathsf{CC}= [\mathbf{p}_0^\mathsf{CC}, \mathbf{p}_1^\mathsf{CC},\ldots,\mathbf{p}_{K}^\mathsf{CC}]$, where both the precoding vector $\mathbf{p}_k^\mathsf{CC}$ for each private multicast message and $\mathbf{p}_0^\mathsf{CC}$ for the super-common message are of size $M \times 1$. 

\item[Mix Coding Scheme ($\mathsf{MC}$):] 
Finally, in the third scheme, the input data vector is defined as $\mathbf{s}^\mathsf{MC} = {[s_0,s_{p,1} + s_0,\ldots,s_{p,K}+s_0]}^T$ $\in \mathbb{C}^{K+1}$. We assume $s_0$ and all $s_{p,k}$ are independent and $\mathbb{E}\{s_0s_0^{\ast}\} = \alpha$ and  $\mathbb{E}\{s_{p,k}s_{p,k}^{\ast}\} = \bar{\alpha}$. The input data vector $\mathbf{s}^\mathsf{MC}$ is linearly processed by a precoder matrix $\mathbf{P}^\mathsf{MC} = [\mathbf{p}_0^\mathsf{MC}, \mathbf{p}_1^\mathsf{MC},\ldots,\mathbf{p}_{K}^\mathsf{MC}]$, where both the precoding vector $\mathbf{p}_k^\mathsf{MC}$ for each multicast data and $\mathbf{p}_0^\mathsf{MC}$ for super-common data are of size $M \times 1$. 
\end{description}
Then, for the transmission schemes $\mathsf{m} \,{\in}\, \{\mathsf{SC,CC,MC}\}$ the overall transmit data vector $\mathbf{x} \in \mathbb{C}^{M }$ can be written as
\begin{align}
\mathbf{x}^\mathsf{m} &= \mathbf{P}^\mathsf{m} \mathbf{s}^\mathsf{m}= \mathbf{p}_A^\mathsf{m} s_0 + \sum_{k \in \mathcal{K}}\mathbf{p}_k^\mathsf{m} s_{p,k}.\label{precodedSignal}
\end{align}
Here, $\mathbf{p}_A^\mathsf{SC}= \sum_{k \in \mathcal{K}} \mathbf{p}_k^\mathsf{SC}$, $\mathbf{p}_A^\mathsf{CC}  = \mathbf{p}_0^\mathsf{CC} $ and $\mathbf{p}_A^\mathsf{MC}  = \mathbf{p}_0^\mathsf{MC}+ \sum_{k \in \mathcal{K}} \mathbf{p}_k^\mathsf{MC}$. Then, the average total power constraint at the BS, $\mathbb{E}\{{\mathbf{x}}^{H}{\mathbf{x}}\}$, can be written as
\begin{align}
   B^\mathsf{m} \|\mathbf{p}_A^\mathsf{m} \|^2+ C^\mathsf{m} \sum_{k \in \mathcal{K}} \|\mathbf{p}_k^\mathsf{m} \|^2\leq E_{tx}. \label{pow_const}
\end{align} In (\ref{pow_const}), $(B^\mathsf{SC} ,C^\mathsf{SC}) =(\alpha, \bar{\alpha} )$, $(B^\mathsf{CC} ,C^\mathsf{CC} ) = (1,1)$ and $(B^\mathsf{MC},C^\mathsf{MC}) =(\alpha, \bar{\alpha} )$. The received signal for the $\mathsf{m}$-th scheme at user-$n$ can be expressed as
\begin{align}
 y_{n}^\mathsf{m}&=\mathbf{h}_n^H \mathbf{p}_A^\mathsf{m} s_0 +  \sum_{k \in \mathcal{K}} \mathbf{h}_n^H \mathbf{p}_k^\mathsf{m} s_{p,k} + z_{n}.\label{rec_signal}
\end{align}In (\ref{rec_signal}), ${\mathbf{h}_n}$ $\in \mathbb{C}^{M}$ is the channel gain vector of the $n$-th user. The entries in $\mathbf{h}_n$ and the noise component $z_n$ are assumed to be independent and identically distributed (i.i.d.). We assume perfect channel state information at the transmitter and the receivers (CSITR).

\subsection{Achievable Data Rates}
In this system, all users decode the super-common message. In addition to this, each user subtracts this super-common message from its received signal to decode its private multicast message using SIC. Then, the achievable rate for super-common and private multicast messages for the $n$-th user for scheme $\mathsf{m}$ are respectively defined as $R_{0,n}^\mathsf{m}$ and $R_{p,n}^\mathsf{m}$ are given as
\begin{align}
R_{0,n}^\mathsf{m}&=  \log \left( 1 + B^\mathsf{m} r_{0,n}^{\mathsf{m}^{-1}}  \left| \mathbf{h}_n^H \mathbf{p}_A^\mathsf{m}  \right |^2 \right),\label{rate_cu}\\
R_{p,n}^\mathsf{m} &= \log \left( 1 + C^\mathsf{m} r_{p,n}^{\mathsf{m}^{-1}} \left|   \mathbf{h}_n^H \mathbf{p}_{\mu(n)}^\mathsf{m}\right|^2\right). \label{rate_u}
\end{align}Here $r_{0,n}^\mathsf{m}$ and $r_{p,n}^\mathsf{m}$  are the effective noise variances for the super-common and private multicast data at the $n$-th user for scheme $\mathsf{m}$. They can be calculated as
\begin{align}
r_{0,n}^\mathsf{m}&= \sum_{k \in \mathcal{K}} C^\mathsf{m}  \left|\mathbf{h}_n^H \mathbf{p}_k^\mathsf{m} \right|^2+ 1,\label{r_0n}
\end{align}
\begin{align}
r_{p,n}^\mathsf{m} &=  r_{0,n}^\mathsf{m} -  C^\mathsf{m}  \left | \mathbf{h}_n^H \mathbf{p}_{ \mu(n)}^\mathsf{m} \right|^2.\label{r_un}
\end{align}Then, the overall achievable rate for the super-common message is determined by the minimum of all $R_{0,n}^\mathsf{m}, n \in \mathcal{N}$, and the achievable rate for private multicast message for group $k$, $s_{p,k}$, is determined by the minimum of all $R_{p,n}^\mathsf{m}, n \in \mathcal{G}_k$. Thus, we also define 
\begin{align}
R_0^\mathsf{m} &= \min_{n \in \mathcal{N}} R_{0,n}^\mathsf{m} , \label{eqn:minRc}\\
R_k^\mathsf{m} &= \min_{n \in \mathcal{G}_k} R_{p,n}^\mathsf{m} . \label{eqn:minRu}
\end{align} 
Note that, the achievable rate for the super-common message, $R_0^\mathsf{m}$, can also be written as a sum of all common rates; i.e. the rate of $S_c$ and $S_{c,k}$, $\forall k$, as follows:
\vspace{-0.5cm}
\begin{align}
R_c^\mathsf{m}+ \sum_{k \in \mathcal{K}} R_{c,k}^\mathsf{m}= R_0^\mathsf{m}. \label{eqn:partialRc}
\end{align}

\subsection{MSE Expressions}
In this subsection, we will utilize the one-to-one correspondence between mutual information and minimum mean square error \cite{guo2005mutual,Christensen2008_WeightedSumRate} to express the rates in  (\ref{rate_cu}) and (\ref{rate_u}) in terms of the MMSE values.  

For MSE estimation, the $n$-th user first processes its received signal $y_n^\mathsf{m}$, with the super-common data receiver $W_n^\mathsf{m}$ to form an estimate of $s_0$, denoted as $\hat{s}_{0,n}= W_n^\mathsf{m}y_n^\mathsf{m}$. Assuming perfect successive interference cancellation, in the second stage, the $n$-th user forms an estimate for the private multicast message $s_{p,k}$ as $\hat{s}_{p,k} = V_n^\mathsf{m} \left(y_n^\mathsf{m} - \mathbf{h}_n^H  \mathbf{p}_{A}^\mathsf{m} s_0  \right)$.

The MSE expressions of super-common and private multicast data for the $n$-th user for scheme $\mathsf{m}$ are respectively defined as
$ \varepsilon_{0,n}^\mathsf{m} = \mathbb{E}\left\{\norm{\hat{s}_{0,n} - s_0}^2 \right\}$, and
$\varepsilon_{p,n}^\mathsf{m}  = \mathbb{E}\left\{\norm{\hat{s}_{p,n} - s_{p,\mu(n)}}^2 \right \}$, and for perfect SIC, their closed form expressions can be written as
\begin{align}
\varepsilon_{0,n}^\mathsf{m} 
&= \left| W_n^\mathsf{m} \right|^2 \left( B^\mathsf{m} | \mathbf{h}_n^H  \mathbf{p}_A^\mathsf{m} |^2 + \sum_{k \in \mathcal{K}}  C^\mathsf{m} | \mathbf{h}_n^H  \mathbf{p}_k^\mathsf{m} |^2 + 1 \right) \nonumber \\
&\quad \:- 2 \mathfrak{R} \left\{ B^\mathsf{m} W_n^\mathsf{m}  \mathbf{h}_n^H  \mathbf{p}_A^\mathsf{m}  \right\}+ B^\mathsf{m} , \label{CMSE_u}\\
\varepsilon_{p,n}^\mathsf{m}
&= \left | V_n^\mathsf{m} \right |^2\left(  \sum_{k \in \mathcal{K}} C^\mathsf{m} |\mathbf{h}_n^H  \mathbf{p}_k^\mathsf{m} |^2 + 1 \right) \nonumber \\
& \quad \: - 2 \mathfrak{R} \left\{C^\mathsf{m} V_n^\mathsf{m} \mathbf{h}_n^H  \mathbf{p}_{\mu(n)}^\mathsf{m}\right\} + C^\mathsf{m}\label{UMSE_k}.
\end{align}When these MSE values attain their minimum, the corresponding receivers are called the optimal MMSE receivers and are defined as $W_n^{\mathsf{m},opt}= \arg\min_{W_n} \varepsilon_{0,n}^\mathsf{m}$ and $V_n^{\mathsf{m},opt} = \arg\min_{V_n} \varepsilon_{p,n}^\mathsf{m}$.
The closed form expressions for these MMSE receivers are then calculated as
\begin{align}
W_n^{\mathsf{m},opt} &= B^\mathsf{m} \mathbf{p}_A^{\mathsf{m}^H} \mathbf{h}_n \left( B^\mathsf{m} \left | \mathbf{h}_{n}^H \mathbf{p}_A^\mathsf{m} \right|^2  + r_{0,n}^\mathsf{m} \right)^{-1},\label{W_rec}\\
V_n^{\mathsf{m},opt} &= C^\mathsf{m} \mathbf{p}_{\mu(n)}^{(m)^H} \mathbf{h}_n r_{0,n}^{\mathsf{m}^{-1}}. \label{V_rec_sic}
\end{align}Given that these MMSE receivers in (\ref{W_rec}) and (\ref{V_rec_sic}) are employed, the resulting error variance expressions in (\ref{CMSE_u}) and (\ref{UMSE_k}) become

\begin{align}
\varepsilon_{0,n}^{\mathsf{m},min}  &=\left( \frac{1}{B^\mathsf{m}} + r_{0,n}^{\mathsf{m}^{-1}}  \left | \mathbf{h}_{n}^H \mathbf{p}_A^\mathsf{m} \right|^2  \right)^{-1}, \label{error_cov_common}\\
\varepsilon_{p,n}^{\mathsf{m},min}  &=\left( \frac{1}{C^\mathsf{m}}  + r_{p,n}^{\mathsf{m}^{-1}} \left |  \mathbf{h}_{n}^H \mathbf{p}_{\mu(n)}^\mathsf{m} \right|^2 \right)^{-1}.\label{error_cov_uni}
\end{align}Comparing (\ref{rate_cu}) and (\ref{rate_u}) with (\ref{error_cov_common}) and (\ref{error_cov_uni}) we can write
\begin{align}
R_{0,n}^\mathsf{m}&= -\log \left( \varepsilon_{0,n}^{\mathsf{m},min}B^{\mathsf{m}^{-1}}\right), \label{rate_cu_errrorCov}\\
R_{p,n}^\mathsf{m} &= -\log\left(\varepsilon_{p,n}^{\mathsf{m},min} C^{\mathsf{m}^{-1}}\right). \label{rate_u_errorCov}
\end{align}

\section{Problem Definition}\label{sec:Probdef}

In this section, we define an optimization problem which aims to find the optimal precoders $\mathbf{P}^\mathsf{m}$ such that the minimum of all cluster rates is maximized subject to a total power constraint and a minimum rate constraint for the common rate. This RS MMF problem is defined as
\begin{IEEEeqnarray}{rl}
\arg\max_{\mathbf{P}^\mathsf{m}, \mathbf{R}_c^\mathsf{m}} &\qquad  \min_{k \in \mathcal{K}} \left(R_{c,k}^\mathsf{m} + \min_{i \in \mathcal{G}_k} R_{p,i}^\mathsf{m}\right)\label{max_SR1}\\
\text{s.t. }&\qquad  R_c^\mathsf{m} + \sum_{k \in \mathcal{K}} R_{c,k}^\mathsf{m} \leq R_{0,n}^\mathsf{m},  \forall n \in \mathcal{N}, \IEEEyessubnumber\label{max_SR1_Rck1}\\
&\qquad  0 \leq R_{c,k}^\mathsf{m}, \forall k \in \mathcal{K},\IEEEyessubnumber \label{max_SR1_Rck2}\\
&\qquad  R_c^{th} \leq R_c^\mathsf{m} \mbox{~and~}  \eqref{pow_const}\IEEEyessubnumber\label{max_SR1_comb2}
\end{IEEEeqnarray}where $\mathbf{R}_c^\mathsf{m} = [ R_c^\mathsf{m}, R_{c,1}^\mathsf{m}, \ldots, R_{c,K}^\mathsf{m}]$ and $R_c^{th}$ is the threshold rate constraint on the common message. We now convert this problem into an equivalent problem as
\begin{IEEEeqnarray}{rl}
\arg\max_{\mathbf{P}^\mathsf{m}, \mathbf{R}_c^\mathsf{m}, \mathbf{R}_k^\mathsf{m} , R_g^\mathsf{m}} &\quad   R_g^\mathsf{m}\label{max_SR1_3}\\
\text{s.t. }&\quad  R_g^\mathsf{m} \leq  R_{c,k}^\mathsf{m} + R_k^\mathsf{m}, \forall k \in \mathcal{K},\IEEEyessubnumber \\
&\quad  R_k^\mathsf{m} \leq R_{p,i}^\mathsf{m}, \forall i \in \mathcal{G}_k, \forall k \in \mathcal{K}, \IEEEyessubnumber \\
&\quad  \eqref{max_SR1_Rck1},  \eqref{max_SR1_Rck2} \mbox{~and~} \eqref{max_SR1_comb2} \IEEEyessubnumber.
\end{IEEEeqnarray}where $\mathbf{R}_k^\mathsf{m} = [ R_1^\mathsf{m}, \ldots, R_K^\mathsf{m}]$, and $R_g^\mathsf{m}$ and $R_k^\mathsf{m}$ are introduced as auxiliary variables to convert the problem. 

The MMF problem defined in \eqref{max_SR1_3} is non-convex due to the non-convex rate expressions. We solve this problem in an iterative fashion, utilizing the relation between mutual information (rate) and MMSE. To do that, we introduce the augmented weighted MSEs (WMSE) defined for the $n$-th user for the $\mathsf{m}$-th signal model as:
\begin{align}
\xi_{0,n}^\mathsf{m}&= w_n^\mathsf{m} \varepsilon_{0,n}^\mathsf{m} - \log_2 (B^\mathsf{m} w_n^\mathsf{m} ) ,\label{ksi_0}  \\
\xi_{p,n}^\mathsf{m} &= v_n^\mathsf{m} \varepsilon_{p,n}^\mathsf{m} - \log_2 (C^\mathsf{m} v_n^\mathsf{m} ) ,\label{ksi_u}
\end{align}where $w_n^\mathsf{m} , v_n^\mathsf{m}  > 0$ are the corresponding weights. Then, the minimum of the augmented WMSEs, defined as
\begin{align}
\xi_{0,n}^{\mathsf{m},min} &\triangleq \min_{w_n^\mathsf{m}, W_n^\mathsf{m}} \xi_{0,n}^\mathsf{m}, \label{ksi_0_MMSE}\\
\xi_{p,n}^{\mathsf{m},min} &\triangleq \min_{v_n^\mathsf{m}, V_n^\mathsf{m}} \xi_{p,n}^\mathsf{m} \label{ksi_u_MMSE}
\end{align}
can be proved to be related with the rate expressions $R_{0,n}^\mathsf{m}$ and $R_{p,n}^\mathsf{m}$ as
\begin{align}
\xi_{0,n}^{\mathsf{m},min} &= 1 - R_{0,n}^\mathsf{m}, \label{ksi_0_MMSE2}
\end{align}
\begin{align}
\xi_{p,n}^{\mathsf{m},min} &= 1 - R_{p,n}^\mathsf{m}. \label{ksi_u_MMSE2}
\end{align}
This result is obtained by checking the first order optimality conditions. By closely examining each augmented WMSE, it can be seen that $\xi_{0,n}^\mathsf{m}$ and $\xi_{p,n}^\mathsf{m}$ are respectively convex in $W_n^\mathsf{m}$ and $V_n^\mathsf{m}$. Then, the optimum receivers in \eqref{ksi_0_MMSE} and \eqref{ksi_u_MMSE} can be found as $W_n^{\mathsf{m} \star} = W_n^{\mathsf{m},opt}$ of \eqref{W_rec}, and $V_n^{\mathsf{m} \star} = V_n^{\mathsf{m},opt}$ of \eqref{V_rec_sic}, and the optimum weights are found as
\begin{align}
w_n^{\mathsf{m}\star} &= w_n^{\mathsf{m},min} =1/\varepsilon_{0,n}^{\mathsf{m},min} , \label{w_weight}\\
v_n^{\mathsf{m} \star} &= v_n^{\mathsf{m},min} = 1/\varepsilon_{p,n}^{\mathsf{m},min}, \label{v_weight}
\end{align}where $\varepsilon_{0,n}^{\mathsf{m},min}$ and $\varepsilon_{p,n}^{\mathsf{m},min}$ are respectively given in \eqref{error_cov_common} and \eqref{error_cov_uni}.


\subsection{Equivalent WMSE Problem}
Motivated by (\ref{ksi_0_MMSE2}) and (\ref{ksi_u_MMSE2}), an equivalent WMSE reformulation for problem (\ref{max_SR1_3}) can be written as:
\begin{IEEEeqnarray}{rl}\label{MSE_all}
\max_{\substack{ \mathbf{P}^\mathsf{m}, \mathbf{R}_c^\mathsf{m}, \mathbf{R}_k^\mathsf{m} , R_g^\mathsf{m},\\ \mathbf{W}^\mathsf{m}, \mathbf{V}^\mathsf{m}, \mathbf{w}^\mathsf{m}, \mathbf{v}^\mathsf{m}}} &\quad   R_g^\mathsf{m}\label{min_MSE1} \\
\text{s.t. }&\quad  R_g^\mathsf{m} \leq  R_{c,k}^\mathsf{m} + R_k^\mathsf{m}, \forall k \in \mathcal{K},\IEEEyessubnumber \\
& \quad R_k^\mathsf{m}  \leq 1 - \xi_{p,i}^\mathsf{m} , \forall i \in \mathcal{G}_k, \forall k \in \mathcal{K}, \IEEEyessubnumber\\
&\quad   R_c^\mathsf{m} + \sum_{k \in \mathcal{K}} R_{c,k}^\mathsf{m} \leq 1 - \xi_{0,n}^\mathsf{m},  \forall n \in \mathcal{N},\label{min_MSE1_Rn0}\IEEEyessubnumber\\
&\quad \eqref{max_SR1_Rck2} \mbox{~and~} \eqref{max_SR1_comb2}.
\end{IEEEeqnarray}In problem \eqref{MSE_all}, $\mathbf{W}^\mathsf{m} = [W_1^\mathsf{m},\ldots, W_N^\mathsf{m} ]$, $\mathbf{V}^\mathsf{m} = [V_1^\mathsf{m},\ldots, V_N^\mathsf{m}]$,  $\mathbf{w}^\mathsf{m} = [w_1^\mathsf{m},\ldots, w_N^\mathsf{m}]$ and $\mathbf{v}^\mathsf{m} = $ $[v_1^\mathsf{m},\ldots, v_N^\mathsf{m}]$. 
The WMSE problem in \eqref{MSE_all} is also a non-convex problem. We solve this problem via an alternating optimization (AO) algorithm given in Algorithm \ref{algorithm:ratesplitting}. In Algorithm \ref{algorithm:ratesplitting}, in each iteration, the receivers $W^\mathsf{m}, V^\mathsf{m}$ and the weights $w^\mathsf{m}, v^\mathsf{m}$ are updated for a given precoder. Afterwards, the precoder $\mathbf{P}^\mathsf{m}$ is updated by solving the problem in \eqref{MSE_all} for the given, newly found receivers and weights. Note that, in each iteration, the problem is convex when the receivers and weights are fixed. Moreover, the algorithm converges to a local optimum. The upper bound on the total power constraint, limits the objective function from above. Since the objective function increases in each iteration of the algorithm, it converges to a local optimum. 

\begin{algorithm}[!t]
    \caption{Proposed WMSE Based Algorithm}
    \label{algorithm:ratesplitting}
    \begin{algorithmic}[1]
        \State \textbf{Initialize:} $\alpha$, $\epsilon$, $E_{tx}$,
        $\textbf{P}^\mathsf{m}$, $\mathsf{R}_\mathsf{c}^\mathsf{th}$, 
        $n = 1$, $R_g^{(0)}, R_g^{({-}1)} \gets 0$
        \While{$\big| R_g^{(n{-}1)} - R_g^{(n{-}2)}
        \big| > \epsilon$} 
            \State Compute $W_n^\mathsf{m}$ and $V_n^\mathsf{m}$ by \eqref{W_rec} and \eqref{V_rec_sic} for given $\textbf{P}^\mathsf{m}$
            \State Compute $\varepsilon_{0,n}^\mathsf{m}$ and $\varepsilon_{p,n}^\mathsf{m}$ by \eqref{CMSE_u} and \eqref{UMSE_k} for given $\textbf{P}^\mathsf{m}$
            \State Compute $w_n^{\mathsf{m},opt}$ and $v_n^{\mathsf{m},opt}$ by \eqref{w_weight} and \eqref{v_weight}
            \State Update $\textbf{P}^\mathsf{m}$ solving \eqref{MSE_all} for given $W_n^\mathsf{m}$,  $V_n^\mathsf{m}$, $w_n^\mathsf{m}$,  $v_n^\mathsf{m}$ \label{algline:update_P}
            \State $R_g^{(n)} \gets $ output of the optimization (\ref{MSE_all})
            \State $n \gets n+1$
        \EndWhile 
    \end{algorithmic}
\end{algorithm} 

\section{Simulation Results}\label{sec:sim}
\begin{figure}[!h]
\centering
\includegraphics[width=5.5in]{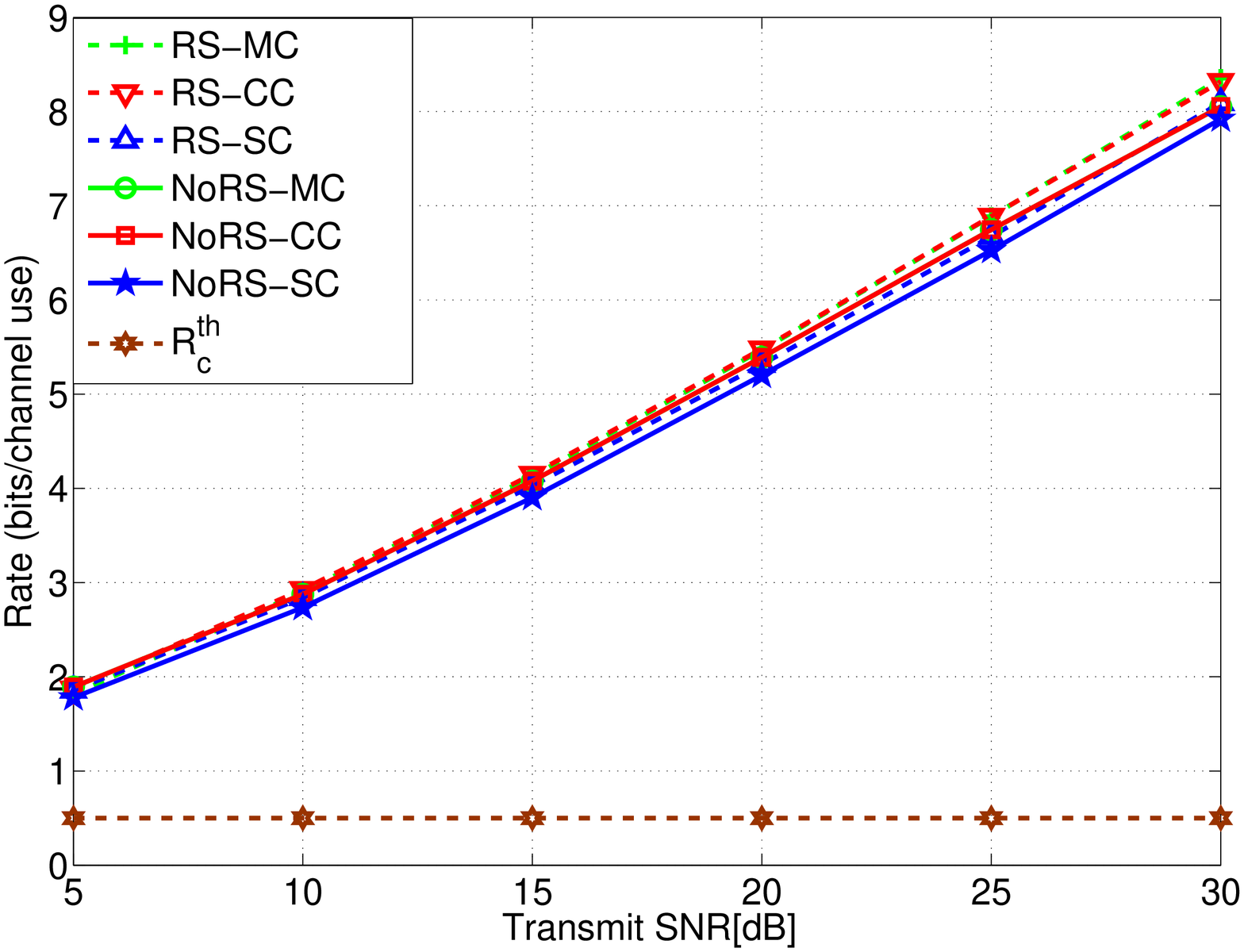}
\caption{MMF Rate for $M=6, N=6, K=[1,2,3], R_c^{th}=[0.5]$. Optimal $\alpha$ is chosen for SC and MC schemes.}\label{fig:M6}
\end{figure}
\begin{figure}[!h]
\centering
\includegraphics[width=5.5in]{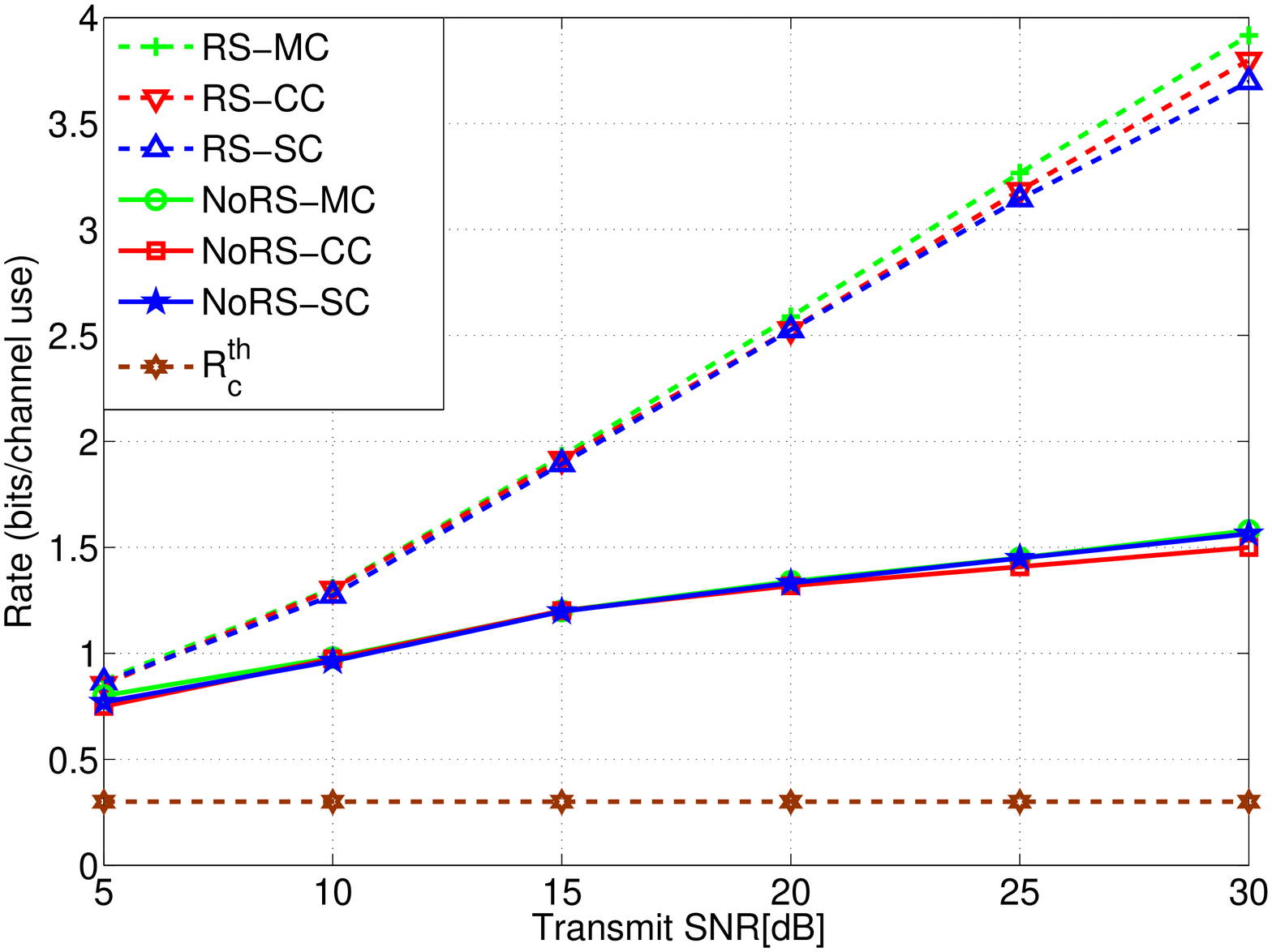}
\caption{MMF Rate for $M=2, N=6, K=[1,2,3], R_c^{th}=[0.3]$. Optimal $\alpha$ is chosen for SC and MC schemes.}\label{fig:M2}
\end{figure}
In this section we provide simulation results to compare the three different precoding schemes. To emphasize that RS is employed in the precoding schemes, in the figures we denote these precoders as $\mathsf{RS-SC}$, $\mathsf{RS-CC}$ and $\mathsf{RS-MC}$ for signal models $\mathsf{SC}$, $\mathsf{CC}$ and $\mathsf{MC}$ respectively. We also compare these schemes with their counterparts with no RS and denote them as $\mathsf{noRS-m}$, $\mathsf{m}= \mathsf{SC,CC,MC}$. In the simulations, the entries in $\mathbf{h}_n$ are assumed to be circularly symmetric complex Gaussian distributed random variables with zero mean and unit variance, and are i.i.d.. Similarly, the noise components $z_n$, $n= 1,...,N$ are i.i.d. circularly symmetric complex Gaussian random variables with zero mean and unit variance. The presented results are averaged over 100 channel realizations. Ideal Gaussian codebooks are assumed. To solve \eqref{MSE_all} in Algorithm \ref{algorithm:ratesplitting}, CVX toolbox \cite{CVX} is used. In addition, we define the transmit SNR as $E_\mathsf{tx}$.


In Figs. \ref{fig:M6} and \ref{fig:M2}, we consider a multi-group multicasting system respectively with $M=6$ and $M=2$ transmit antennas in which $N=6$ users are split into $K=3$ groups with $G_1 = 1$, $G_2 = 2$ and $G_3 = 3$ users in each group. In both figures, $R_c^{th}$ is satisfied for all channel realizations, and the system-wide common message rate is equal to $R_c^{th}$. 

The system in  Fig. \ref{fig:M6} represents an underloaded system. It is observed that the proposed RS schemes and no RS schemes with a system-wide common message have almost the same performance. On the other hand, Fig. \ref{fig:M2} shows the results for an overloaded multi-group multicasting system. In overloaded systems, interference management is crucial since the number of transmit antennas is insufficient to serve all the users. In this figure, it is shown that RS is essential for managing inter-group interference. There is a significant difference between RS and NoRS schemes. While NoRS schemes saturate in the interference limited region (high SNR), the MMF rate for RS schemes keeps increasing. Results confirm that observations of \cite{Joudeh2017_RateSplittingforMaxMin} on the usefulness of RS in an overloaded multi-group multicast also carries on to the case where a common message is additionally transmitted.

Fig. \ref{fig:M2} also shows that when RS schemes are compared with each other in terms of their MMF rates, the MC scheme is the best, and CC performs better than SC. However, when they are compared in terms of their complexities, CC scheme is the best. This is because the optimal $\alpha$ has to be found for MC and SC, while CC does not depend on the $\alpha$ parameter. 

\section{Conclusion}\label{sec:conc}

This letter investigates precoding for maximizing the minimum of all cluster rates in a multi-group multicasting system with a system-wide common message. Three different precoding schemes based on rate-splitting are suggested and an alternating optimization procedure is proposed to solve for the maximally fair cluster rate. The proposed schemes are compared with their counterparts with no rate-splitting. Simulation results show that rate-splitting is an essential precoding technique  especially in overloaded systems. 

\ifCLASSOPTIONcaptionsoff
  \newpage
\fi



%
\bibliographystyle{IEEEtran} 

\bibliography{references}
\end{document}